\documentclass{article}
\title{\bf Spinning fluid cosmology}
\author{Morteza Mohseni\thanks{email:m-mohseni@pnu.ac.ir}
\\{\small Physics Department, Payame Noor University, Tehran 19395-4697,
Iran}}
\begin{document}
\maketitle
\begin{abstract}
\noindent The dynamics of a spinning fluid in a flat cosmological model is investigated.
The space-time is itself generated by the spinning fluid which is characterized by an energy-momentum tensor consisting a sum of the usual perfect-fluid energy-momentum tensor and some Belinfante-Rosenfeld tensors. It is shown that the equations of motion admit a solution for which the fluid four-velocity and four-momentum are not co-linear in general. The momentum and spin densities of the fluid are expressed in terms of the scale factor.


\end{abstract}
\section{Introduction}\label{intro}
The study of the dynamics of fluids has been an essential part of
the general theory of relativity with wide range of applications
in cosmology and certain areas of astrophysics. An interesting
type of fluid is the spinning fluid, i.e., a perfect fluid
consisting of particles carrying an internal but classical spin.
Following Weyssenhoff and Raabe~\cite{wys} in which a model was
proposed for spinning fluids, they have attracted interest both in
the frameworks of the general relativity and the Einstein-Cartan
theories. Ray et al.~\cite{ray,ray2} proposed an energy-momentum tensor
for spinning fluids in general relativity. This was then used
in~\cite{fen,bed} to obtain a spin-included Raychaudhuri equation. A
self-consistent formulation of spinning fluids has been
presented in~\cite{oli}-\cite{mar} by including the spin as a thermodynamic
variable in the theory. In Ref.~\cite{obu} a variational theory of
an ideal fluid with spin was formulated in Riemannian space-time
within the framework of general relativity theory. The fluid
energy-momentum tensor of this model is a modification of the
energy-momentum tensor used in~\cite{ray}. The resulting equations were used there to show that the spin has no effect on the standard cosmology in the sense that the pressure and the energy density obey the same equation as if one uses the perfect fluid model. In these models the
so-called Frenkel condition is used in the course of deriving some
of the equations. The equations of motion obtained in
Ref.~\cite{obu} was generalized in~\cite{yur} by adding an
effective potential term, and by choosing some specific form of
this effective potential, an inflationary cosmological model with
self-interacting spinning matter was suggested.
A spin-dominated inflation model has also been suggested in~\cite{gas}.
Some other cosmological applications of spinning fluids may be found in,
e.g.,~\cite{cos}-\cite{yua}. Another interesting model for spinning fluids was developed in
\cite{bai}. In that model the energy-momentum tensor was derived by variation
of a Lagrangian whose explicit form is not needed to be specified and the
equation of motion are obtained without resort to a supplementary condition.
The relevant equations of motion reduce to the correct form of the Mathisson-Papapetrou-Dixon equations \cite{dix} in the case of spinning dusts. The later
are widely used to describe the motion of spinning particles in curved backgrounds.

The aim of the present work is to investigate the motion of spinning fluids
in an FRW space-time. We are particularly interested in nontrivial solutions to the equations of motion. By nontrivial solutions we mean those solutions which exhibit the non-co-linearity of the four-momentum and four-velocity. In fact it can be shown that the equations of motion admit a trivial solution for which the four-momentum is proportional to the four-velocity (see also \cite{kor}).

To describe the motion of spinning fluid in the space-time generated by it, we deploy the spinning fluid model of Ref.~\cite{bai}. We find that
in this case a consistent treatment of space-time-spinning fluid is possible and generates a flat FRW cosmology. We then obtain the fluid momentum and spin in terms of the scale factor.

This paper is organized as follows. In the following section the spinning fluid energy-momentum tensor is reviewed and the corresponding equations of motion are collected.  Next we consider the motion of the fluid in a flat FRW universe. We solve the equations of motion for the fluid momentum and spin. In the last section we present our conclusions.
\section{Spinning fluids}\label{sec:1}
A general relativistic spinning fluid is described by the following energy-momentum tensor
\begin{equation}\label{e18}
T_{\mu\nu}=h_{\mu\nu}p+p_\mu
u_\nu-\frac{1}{2}\nabla_\alpha({s^\alpha}_\mu
u_\nu+{s^\alpha}_\nu u_\mu+s_{\mu\nu}u^\alpha)
\end{equation}
where $h_{\mu\nu}=g_{\mu\nu}+u_\mu u_\nu$ is a projection tensor with $u^\mu$ being the
fluid four-velocity, $p^\mu$ represents the fluid four-momentum density, $s_{\mu\nu}$ is an antisymmetric tensor corresponding to the fluid spin density, $p$ denotes the fluid pressure, and $\nabla_\alpha$ stands for covariant derivative. This energy-momentum tensor
is a sum of the usual perfect-fluid energy-momentum tensor and a term containing the Belinfante-Rosenfeld tensor. This term disappears if the spin is turned off.

The energy-momentum tensor is required to be symmetric. By imposing this requirement on the above tensor we obtain the spin equation of motion
\begin{equation}\label{eq1}
p_\mu u_\nu-p_\nu u_\mu=\nabla_\alpha(u^\alpha s_{\mu\nu})
\end{equation}
which shows that the spinning fluid four-velocity and four-momentum are not co-linear in general. This equation may be used to recast the energy-momentum tensor into the following manifestly symmetric form
\begin{equation}\label{e18a}
T_{\mu\nu}=h_{\mu\nu}p+\frac{1}{2}(p_\mu
u_\nu+p_\nu u_\mu)-\frac{1}{2}\nabla_\alpha({s^\alpha}_\mu
u_\nu+{s^\alpha}_\nu u_\mu).
\end{equation}
Translational equations of motion may be obtained either from the action leading to the above energy-momentum tensors by a variation of world-lines \cite{bai}, or more
straightforwardly, by demanding the energy-momentum tensor to satisfy the conservation equation $\nabla_\nu T^{\mu\nu}=0$. For (\ref{e18}), the later results in
\begin{eqnarray}\label{eq2}
h^{\mu\nu}\partial_\nu p+p\nabla_\nu(u^\mu u^\nu)
+\nabla_\nu(p^\mu u^\nu)=-\frac{1}{2}{R^\mu}_{\nu\alpha\beta}u^\nu
s^{\alpha\beta}
\end{eqnarray}
where ${R^\mu}_{\nu\alpha\beta}$ is the Riemann tensor given by
$${R^\mu}_{\nu\alpha\beta}A^{\nu}=(\nabla_\alpha\nabla_\beta-\nabla_\beta\nabla_\alpha)
A^\mu.$$
The above equation of motion may be projected parallel to $u^\mu$
to give
\begin{equation}\label{eq4a}
(p+\rho)\theta+u^\mu\partial_\mu\rho=-p_\mu u^\alpha\nabla_\alpha u^\mu
\end{equation}
in which $\theta=\nabla_\mu u^\mu$ and we have used $u_\mu p^\mu=-\rho$, $\rho$ being
the energy (mass) density. This is a conservation equation.
Similarly projection normal to $u^\mu$ yields
\begin{eqnarray}
h^{\mu\nu}\partial_\nu p+p{\dot u}^\mu+{\dot p}^\mu+(p^\mu-\rho u^\mu)\theta-{\dot\rho}u^\mu-u^\mu{\dot u^\kappa}p_\kappa=-\frac{1}{2}{R^\mu}_{\nu\alpha\beta}u^\nu s^{\alpha\beta}\label{eq4b}
\end{eqnarray}
which is an equation of motion for the fluid momentum density.
\section{The cosmology}\label{sec:2}
In this section we aim to investigate the dynamics of a spinning fluid in a cosmological model. Consider a flat FRW space-time whose line element is given by
\begin{equation}\label{eq23}
ds^2=-dt^2+a^2(t)\delta_{ij}dx^idx^j
\end{equation}
where $a(t)$ is the scale factor and $(t,x^i),\,i=1,2,3$, are the coordinates. For this
space-time we have the following standard geometrical data
\begin{eqnarray*}
\Gamma^0_{ii}&=&a(t){\dot a}(t),\,\Gamma^i_{0i}=\frac{{\dot a}(t)}{a(t)}\\
R_{0i0i}&=&-a(t){\ddot a}(t),\,R_{ijij}=(a(t){\dot a}(t))^2\\
G_{00}&=&-3\left(\frac{\dot a(t)}{a(t)}\right)^2,\,G_{ii}={\dot a(t)}^2+2a(t){\ddot a(t)}
\end{eqnarray*}
In a co-moving frame the four-velocity is $u^\mu=(1,0,0,0)$ and the four-momentum is given by $p^\mu=(\rho(t),p^i)$. Thus we have
\begin{eqnarray}
T^{00}&=&\rho,\label{el1}\\
T^{ii}&=&\frac{p}{a^2},\label{el2}\\
T^{0i}&=&\frac{1}{2}\left(p^i-\partial_t s^{0i}-2\frac{{\dot a}(t)}{a(t)}s^{0i}\right)\label{el3}.
\end{eqnarray}
Inserting these into the Einstein's equation
\begin{eqnarray*}
G_{\mu\nu}=-\kappa T_{\mu\nu},\hspace{3mm}\kappa=\frac{8\pi G}{c^4}
\end{eqnarray*}
we obtain
\begin{eqnarray}
\rho(t)&=&3\kappa^{-1}\left(\frac{\dot a(t)}{a(t)}\right)^2,\label{el1a}\\
p(t)&=&-\kappa^{-1}\left(\frac{\dot a^2(t)}{a^2(t)}+2\frac{\ddot a(t)}{a(t)}\right),\label{el2a}\\
p^i(t)&=&\partial_t s^{0i}+2\frac{{\dot a}(t)}{a(t)}s^{0i}\label{el3a}.
\end{eqnarray}
The first two equations relate the scale factor to the energy density and pressure of the spinning fluid. These are the same as the corresponding equations for a usual perfect fluid. The third equation establishes a connection between the spacial components of the momentum density and the electric components of the spin density tensor.

The spin magnetic components may be obtained from the spin equation of motion (\ref{eq1})
which in this case reads
\begin{equation}\label{el4}
\frac{d}{dt}s^{ij}(t)+5\frac{\dot a(t)}{a(t)}s^{ij}=0.
\end{equation}
This results in
\begin{equation}\label{el4a}
s^{ij}(t)=l^{ij}(a(t))^{-5}
\end{equation}
$l^{ij}$ being integration constants. Similarly we can obtain the electric components
by solving the corresponding equation
\begin{equation}\label{el4b}
\frac{d}{dt}s^{0i}(t)+4\frac{\dot a(t)}{a(t)}s^{0i}=-p^i(t)
\end{equation}
for $s^{0i}$. Combining this with equation (\ref{el3a}) we obtain
\begin{eqnarray}
s^{0i}&=&l^{0i}(a(t))^{-3},\label{el5}\\
p^i(t)&=&-l^{0i}{\dot a(t)}(a(t))^{-4}\label{el5a}
\end{eqnarray}
with $l^{0i}$ being constants. On the other hand, the components of momentum can be obtained directly from the translational equation of motion. The reduced form of equation
(\ref{eq4b}) is
\begin{equation}\label{el6}
\frac{d}{dt}p^i(t)+4\frac{\dot a(t)}{a(t)}p^i=-\frac{\ddot a(t)}{a(t)}\,s^{0i}.
\end{equation}
This is compatible with equations (\ref{el5}) and (\ref{el5a}).

From the above expressions for the fluid momentum and spin densities, it is possible
to obtain the total momentum and spin contained in a spacial volume $V=\int\sqrt{-g}d^3x$.
We have
\begin{eqnarray}
\int\sqrt{-g}s^{0i}d^3x&=&l^{0i}V,\label{el50}\\
\int\sqrt{-g}s^{ij}d^3x&=&l^{ij}V(a(t))^{-2},\label{el51}\\
\int\sqrt{-g}p^i(t)d^3x&=&-l^{0i}V H\label{el52}
\end{eqnarray}
where $$H=\frac{\dot a(t)}{a(t)}.$$
\section{Conclusions}\label{sec:3}
We have obtained equations describing the motion of spinning fluid in a flat FRW space-time which is itself generated by the fluid. The equations of motion are self-sufficient
and no supplementary condition is used. These equations result in a general solution
describing a spinning fluid whose momentum and velocity four-vectors are not co-linear in general. To our knowledge, this is the first solution of this kind for spinning fluids. For a particular set of integration constants the momentum and velocity become co-linear and our results are in complete
agreement with those of ref. \cite{obu} in this case. The existence of this later special solution might be related to the maximal symmetries
of the cosmological model under consideration. The evolution of the scale factor is independent of spin in this model. The total electric components of the fluid spin is constant but the magnetic components are proportional to the inverse square of the scale factor. The total momentum is proportional to the Hubble parameter.


\begin{thebibliography}{}
\bibitem{wys} Weyssenhoff, J., Raabe, A., Acta. Phys. Pol. \textbf{9}, 7 (1947).
\bibitem{ray} Ray, J.R., Smalley, L.L., Phys. Rev. D \textbf{26}, 2619 (1982).
\bibitem{ray2} Ray, J.R., Smalley, L.L., Phys. Rev. D \textbf{27}, 1383 (1983).
\bibitem{fen} Fennelly, A.J., Krisch, J.P., Ray, J.R., Smalley, L.L., J. Math. Phys. \textbf{32}, 485 (1991).
\bibitem{bed} Bedran, M.L., Vasconcellos-Vaidya, E.P., Lett. Nuovo Cimento, \textbf{41}, 73  (1984).
\bibitem{oli} Ray, J.R., Smalley, L.L., Krisch, J.P., Phys. Rev. D \textbf{35}, 3261 (1987).
\bibitem{sal} de Oliveira, H.P., Salim, J.M., Class. Quantum Grav. \textbf{8}, 1199 (1991).
\bibitem{mar} Martins, M.A., Vasconcellos-Vaidya, E.P., Som, M.M., Class. Quanrum Grav.
\textbf{8}, 2225 (1991).
\bibitem{obu} Obukhov, Y.N., Piskareva, O.B., Class. Quantum Grav. \textbf{6}, L15 (1989).
\bibitem{yur} Obukhov, Y.N., Phys. Lett. A \textbf{182}, 214 (1993).
\bibitem{gas} Gasperini, M., Phys. Rev. Lett. \textbf{56}, 2873 (1986).
\bibitem{cos} Som, M.M., Bedran, M.L., Vasconcellos-vaidya, E.P., Phys. Lett. A \textbf{117}, 169 (1986).
\bibitem{ber} Berman, M.S., Nuovo Cimento B \textbf{105}, 235 (1990).
\bibitem{yua} Yuanjie, L., Int. J. Theor. Phys. \textbf{32}, 667 (1993).
\bibitem{bai} Bailey, I., Ann. Phys. \textbf{119}, 76 (1979).
\bibitem{dix} Dixon, W.G., Proc. R. Soc. London, ser. A \textbf{314}, 499 (1970).
\bibitem{kor} Obukhov, Y.N., Korotky, V.A., Class. Quantum Grav. \textbf{4}, 1633 (1987).
\end{thebibliography}
\end{document}